\documentclass[prb,aps,twocolumn,superscriptaddress,showpacs]{revtex4-1}
\usepackage[colorlinks=true,linkcolor=black, citecolor=blue, urlcolor=blue, 
    unicode=true]{hyperref}

\usepackage{graphicx}
\usepackage{epstopdf}

\usepackage{color}

\definecolor{model1}{RGB}{0,0,0}
\definecolor{model2}{RGB}{255,0,0}
\definecolor{model3}{RGB}{135,81,81}
\definecolor{model4}{RGB}{0,176,80}
\definecolor{model5}{RGB}{126,49,123}
\definecolor{model6}{RGB}{128,128,0}
\definecolor{model7}{RGB}{0,0,255}

\begin{document}

\title{Relaxing Kondo screened Kramers-doublets in CeRhSi$_{3}$}

\author{J. P\'{a}sztorov\'{a}}
\author{A. Howell}
\author{M. Songvilay}
\affiliation{School of Physics and Astronomy, University of Edinburgh, Edinburgh EH9 3JZ, UK}

\author{P.~M.~Sarte}
\affiliation{School of Chemistry, University of Edinburgh, Edinburgh EH9 3FJ, UK}

\author{J. A. Rodriguez-Rivera}
\affiliation{NIST Center for Neutron Research, National Institute of Standards and Technology, 100 Bureau Drive, Gaithersburg, Maryland, 20899, USA}
\affiliation{Department of Materials Science, University of Maryland, College Park, Maryland 20742, USA}

\author{A.~M.~Ar\'{e}valo-L\'{o}pez} 
\affiliation{Univ. Lille, CNRS, Centrale Lille, ENSCL, Univ. Artois, UMR 8181 - UCCS, F-59000 Lille, France}

\author{K. Schmalzl}
\affiliation{Forschungszentrum Juelich GmbH, Juelich Centre for Neutron Science at ILL, 71 avenue des Martyrs, 38000 Grenoble, France}

\author{A. Schneidewind}
\affiliation{Forschungsneutronenquell Heinz Meier-Leibnitz (FRM-II), D-85747 Garching, Germany}
\affiliation{Institut f\"{u}r Festk\"{o}rperphysik, TU Dresden, D-1062 Dresden, Germany}

\author{S. R. Dunsiger}
\affiliation{Physics Department, James Franck Strasse 1, Technical University of Munich, D-85748 Garching, Germany}
\affiliation{Centre for Molecular and Materials Science, TRIUMF, Vancouver, British Columbia V6T 2A3, Canada}

\author{D. K. Singh}
\affiliation{Department of Physics and Astronomy, University of Missouri, Missouri 65211, USA}

\author{C. Petrovic, R. Hu}
\affiliation{Condensed Matter Physics, Brookhaven National Laboratory, Upton, New York,11973, USA}

\author{C. Stock}
\affiliation{School of Physics and Astronomy, University of Edinburgh, Edinburgh EH9 3JZ, UK}

\date{\today}

\begin{abstract}

CeRhSi$_{3}$ is a superconductor under pressure coexisting with a weakly antiferromagnetic phase characterized by a Bragg peak at $\vec{q}_{0}$=($\sim$ 0.2, 0, 0.5) (N. Aso \textit{et al.} J. Magn. Magn. Mater. {\bf{310}}, 602 (2007)).  The compound is also a heavy fermion material with a large specific heat coefficient $\gamma$=110 mJ $\cdot$ mol$^{-1}$ $\cdot$ K$^{-2}$ and a high Kondo temperature of $T_{K}$=50 K indicative that CeRhSi$_{3}$ is in a strongly Kondo screened state.  We apply high resolution neutron spectroscopy to investigate the magnetic fluctuations in the normal phase, at ambient pressures, and at low temperatures.  We measure a commensurate dynamic response centered around the $\vec{Q}$=(0, 0, 2) position that gradually evolves to H~$\sim$ 0.2 with decreasing temperature and/or energy transfers.   The response is broadened both in momentum and energy and not reminiscent of sharp spin wave excitations found in insulating magnets where the electrons are localized.  We parameterize the excitation spectrum and temperature dependence using a heuristic model utilizing the random phase approximation to couple relaxing Ce$^{3+}$ ground state Kramers doublets with a Kondo-like dynamic response.  With a Ruderman-Kittel-Kasuya-Yosida (RKKY) exchange interaction within the $ab$ plane and an increasing single site susceptibility, we can qualitatively reproduce the neutron spectroscopic results in CeRhSi$_{3}$ and namely the trade-off between scattering at commensurate and incommensurate positions.   We suggest that the antiferromagnetic phase in CeRhSi$_{3}$ is driven by weakly correlated relaxing localized Kramers doublets and that CeRhSi$_{3}$ at ambient pressures is on the border between a Rudderman-Kittel-Yosida antiferromagnetic state and a Kondo screened phase where static magnetism is predominately absent.

\end{abstract}

\maketitle

\section{Introduction}

There are a growing list of heavy fermion based materials that show a balance between unconventional superconductivity and localized magnetism.~\cite{Stewart84:56,Pfleiderer09:81}  In particular, several Ce$^{3+}$ compounds, such as CeCoIn$_{5}$ (T$_{c}$=2.3 K), display similar unconventional superconducting order parameters~\cite{Izawa01:87,Mathur98:394} to that of cuprate or iron based high temperature superconductors.  However, in case of the Ce$^{3+}$ materials, the energy scale is much more amenable to experimental techniques such as thermal and cold neutron spectroscopy.~\cite{Stockert11:7,Stock08:100,Stock12:109_2} These materials therefore provide excellent model systems to understand the balance between competing electronic and magnetic phases.~\cite{Coleman87:35,Burdin00:85}  We present neutron inelastic scattering data in the low temperature ambient and non superconducting phase of CeRhSi$_{3}$~\cite{Kawai07:76} illustrating the competition between commensurate magnetism and incommensurate density wave fluctuations.

CeRhSi$_{3}$ is a noncentrosymmetric heavy fermion (space group \# 107 $I4mm$) with a tetragonal unit cell with lattice parameters $a$=4.269 \AA\ and $c$=9.738 \AA.~\cite{Muro98:67} Magnetic Ce$^{3+}$ ions are located at the body center and the unit cell edges.  The body center Ce$^{3+}$ position is surrounded by a layer of Rh above and by a layer of Si below, therefore breaking inversion symmetry making CeRhSi$_{3}$ non centrosymmetric (Fig. \ref{structure_cef} $a$).   CeRhSi$_{3}$ has a comparatively high electronic specific heat coefficient of $\gamma$=110 mJ $\cdot$ mol$^{-1}$ $\cdot$ K$^{-2}$ in comparison to, for example, antiferromagnetic CeCoGe$_{3}$ with $\gamma$=57 mJ $\cdot$ mol$^{-1}$ $\cdot$ K$^{-2}$ along with a large Kondo temperature of T$_{K}$=50 K.~\cite{Kimura07:76}  Putting these values together to estimate the RKKY temperature (given by $k_{B}T_{RKKY}\sim 3 J^{2} \gamma/ \pi^{2}$ with $J$ estimated from the specific heat and Kondo temperature with $J=-1/\log(3 T_{K} \gamma/ \pi^{2})$)~\cite{Yang08:454} gives T$_{RKKY}\sim300$ K.  While T$_{RKKY}$$>$T$_{K}$ is indicative of  local magnetism~\cite{Coleman01:13}, the combination of $\gamma$ and T$_{K}$ is suggestive of a strongly Kondo screened phase by comparing $-1/\log(3 k_{B}T_{K} \gamma \pi^{-2})$ to other Cerium based heavy fermions.~\cite{Yang08:454}  Furthermore,  the low temperature phase has been studied by both band structure calculations and quantum oscillation measurements reporting large electron effective mass enhancement factors of $m^{*}/m_{e} \sim 8$.~\cite{Terashima08:78}   Possibly indicative of being on the borderline between itinerant and localized magnetism, CeRhSi$_{3}$ is weakly antiferromagnetic with a small ordered magnetic moment $\sim$ 0.1 $\mu_{B}$ below temperatures of T$_{N}$=1.6 K and was previously characterized by an incommensurate wave vector of ($\sim$0.2, 0, 0.5) investigated with neutron elastic scattering.~\cite{Aso07:310}

Under pressures greater than $\sim$ 12 kbar, CeRhSi$_{3}$ is found to be superconducting with a T$_{c} \sim$ 1 K~\cite{Kimura05:95,Kimura07:76} and unusually large critical fields along the $c$-axis.~\cite{Kimura11:80,Tada08:101,Kimura07:98,Sugawara10:79}  The superconducting phase is further peculiar as muon spectroscopy finds antiferromagnetism to persist into this phase only being completely suppressed at pressures of 23.6 kbar following a second order transition indicative of a quantum critical point.~\cite{Egetenmeyer12:108}  This is consistent with AC susceptibility measurements under pressure.~\cite{Sugawara11:80}  The nature of this critical point has been further explored  by resistivity measurements under pressure and applied magnetic field~\cite{Iida13:250}, but remains unclear due to the lack of a high pressure and high field Fermi liquid phase.  However, we note that recent penetration depth measurements report magnetism and superconductivity coexisting up to the largest pressures measured.~\cite{Landaeta18:97}  

The noncentrosymmetric crystal structure combined with the strongly correlated electronic nature evidenced by heat capacity, allows the possibility of novel unconventional electronic orders.~\cite{Tanase07:76,Chang17:13}  In centrosymmetric superconductors, spin-up and spin-down bands are degenerate in energy, however when the crystal structure is non centrosymmetric, spin-orbit coupling can split these two bands and the orbital and spin wave functions cannot be treated independently.~\cite{Gorkov01:87}  Magnetic fluctuations may therefore play a key role in superconductivity~\cite{Tada10:81,Tada08:69,Tada08:77} and this has been further implicated by the transport and spectroscopic measurements outlined above which seem to suggest that the Neel and superconducting temperatures are correlated.

We characterize the magnetic fluctuations in the low temperature Kondo screened, normal state, and at ambient conditions in CeRhSi$_{3}$ by applying neutron inelastic scattering.  This paper is divided into six sections including this introduction and a final summary and conclusions component.  We first discuss the single-ion crystal field theory to motivate our spectroscopic experiments.  The experiments and materials are described in the next section along with how we utilized the multidetector array on MACS to correct for the background.  We then present our data illustrating the competing commensurate and incommensurate response in CeRhSi$_{3}$ and discuss the results in terms of a heuristic model based on weakly coupled Kramers doublets. 

\section{Single-ion crystal field theory}

\begin{figure}[t]
\includegraphics[width=8.7cm] {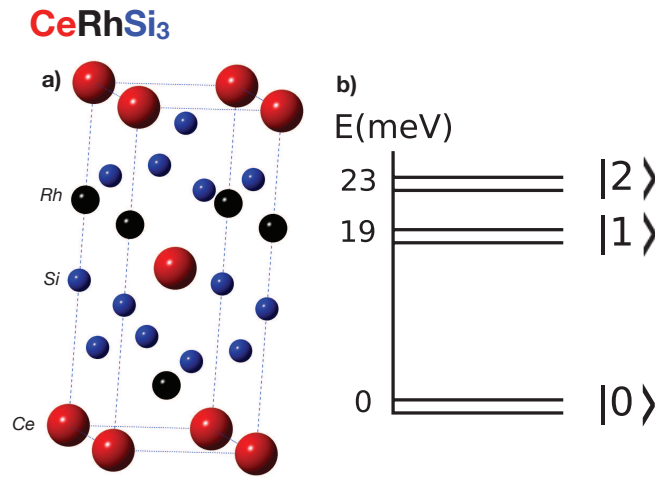}
\caption{\label{structure_cef} $(a)$ The $I4mm$ (No. 107) crystal structure of CeRhSi$_{3}$ and $(b)$ the crystal field scheme discussed in the text using the Steven's parameters extracted by Muro \textit{et al.} in Ref. \onlinecite{Muro07:76}.}
\end{figure}

As a starting point towards understanding the neutron scattering cross section characterizing the magnetic excitations and static order in CeRhSi$_{3}$, we review the crystal field theory results and also calculate matrix elements relevant for the discussion of the static and dynamic magnetism in this heavy fermion compound.  In the discussion section below, we apply this analysis to investigate the possibility that the thermally isolated ground state doublet can be considered in terms of projecting onto a $j_{eff}=1/2$ angular momentum operator.  For Ce$^{3+}$ with $j$=5/2 in a local $C_{4\nu}$ environment, as is the case for CeRhSi$_{3}$ in Fig. \ref{structure_cef} $(a)$, the crystal field scheme should consist of 3 Kramers doublets.  The corresponding Hamiltonian takes the following form,

\begin{equation}
 \mathcal{H}_{CEF} = B_{2} ^{0} O_{2} ^{0} + B_{4} ^{0} O_{4} ^{0} + B_{4} ^{4} O_{4} ^{4}
 \end{equation}
 
 \noindent where $O_{n}^{m}$ are the Stevens operators based on angular momentum operators acting on the $|j,m\rangle$ basis and  $B_{n}^{m}$ are the Stevens parameters.~\cite{Hutchings64:16}
 
To obtain an idea about how much neutron scattering intensity should reside in the different crystal field doublets and in particular the intensity in the inelastic and elastic channels, we have taken the following Stevens parameters extracted from susceptibility measurements.  
 
\begin{table}[ht]
\caption{Stevens coefficients taken from Ref. \onlinecite{Muro07:76}.}
\centering
\begin{tabular} {c c}
\hline\hline
$B_{2}^{0}$= &-0.151 meV \\
$B_{4}^{0}$= & 0.0329 meV \\
$B_{4}^{4}$= & 0.409 meV  \\
\hline
\label{table_poly}
\end{tabular}
\end{table}

\noindent  The resulting crystal field scheme is illustrated in Fig. \ref{structure_cef} $(b)$ and found to be in good agreement with the energy values of the crystal field excitations extacted from neutron inelastic scattering experiments.  Substituting these coefficients into $\mathcal{H}_{CEF}$, we can obtain the eigenstates and calculate the matrix elements characterizing the elastic and inelastic neutron cross sections.  We then obtain the following cross sections for neutron scattering exciting dynamics with the ground-state doublet (denoted as $I_{inelastic}$) and from the elastic cross section ($I_{elastic}$)~\cite{Birgeneau71:4,Turberfield71:42}, 

\begin{eqnarray}
I_{inelastic} \propto \sum_{i=x,y} |\langle -|J^{i}|+\rangle|^{2} \mu_{B}^2=2.3 \mu_{B}^2 \nonumber \\
I_{elastic}\propto g_{J}^2|\langle 0|J^{z}| 0\rangle|^{2} \mu_{B}^{2}\sim 10^{-3} \mu_{B}^2,\nonumber
\end{eqnarray}

\noindent with $g_{J}$=6/7 for Ce$^{3+}$.  The above calculation suggests, based on the single-ion crystal field Hamiltonian, a large dynamic cross section in the neutron scattering response and comparatively little intensity in the elastic channel, a result consistent with initial powder work with neutrons.~\cite{Krimmel02:74}  A similar situation exists with YbRh$_{2}$Si$_{2}$ with a small ordered moment in comparison with the spectral weight in the inelastic channel.~\cite{Stock12:109}  The crystal field prediction of a low static ordered magnetic moment at low temperature is consistent with the small entropy gain measured from heat capacity (only 0.08 $R\ln 2$) which is one of the smallest values found for Ce$^{3+}$ based materials.~\cite{Kimura07:76}  The small static magnetic moment may therefore possibly be indicative of the underlying crystalline electric field rather than quantum criticality~\cite{Si10:329} which would mark the boundary between an itinerant phase and one where the mass of electrons diverge being characterized by more localized magnetism.~\cite{Coleman01:13}  Such a critical point has also been suggested to host a Griffiths phase.~\cite{Neto98:81}  The lack of quantum criticality has been supported by transport measurements under pressure~\cite{Sugawara12:81} and the presence of low-energy spin fluctuations has been implicated as the origin of the linear resistivity.~\cite{Tada10:81}

\begin{figure}[t]
\includegraphics[width=8.8cm] {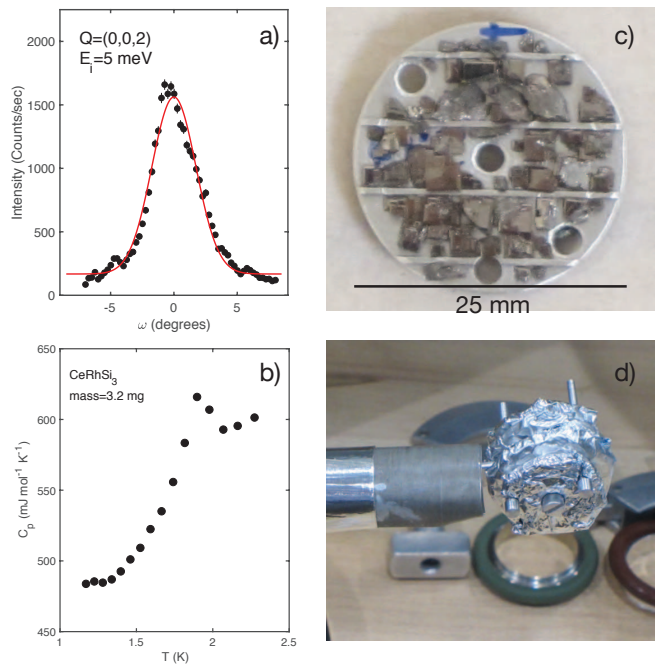}
\caption{\label{sample_mount} $(a)$ The rocking curve through the (002) Bragg peak with a full width at half maximum of 4.0$^{\circ}$ at room temperature.  $(b)$ Heat capacity data as function of temperature for CeRhSi$_{3}$ confirming a transition at $\sim$ 1.8 K. $(c)$-$(d)$ The sample mount used for the neutron inelastic scattering studies on CeRhSi$_{3}$ discussed in the main text.}
\end{figure}
	
\section{Experimental details}

Having outlined the background and crystal field theory describing the magnetic excitations, we now discuss the experimental techniques and results obtained with neutron scattering.

\subsection{Materials and scattering experiments:}

\begin{figure*}[t]
\includegraphics[width=18cm]{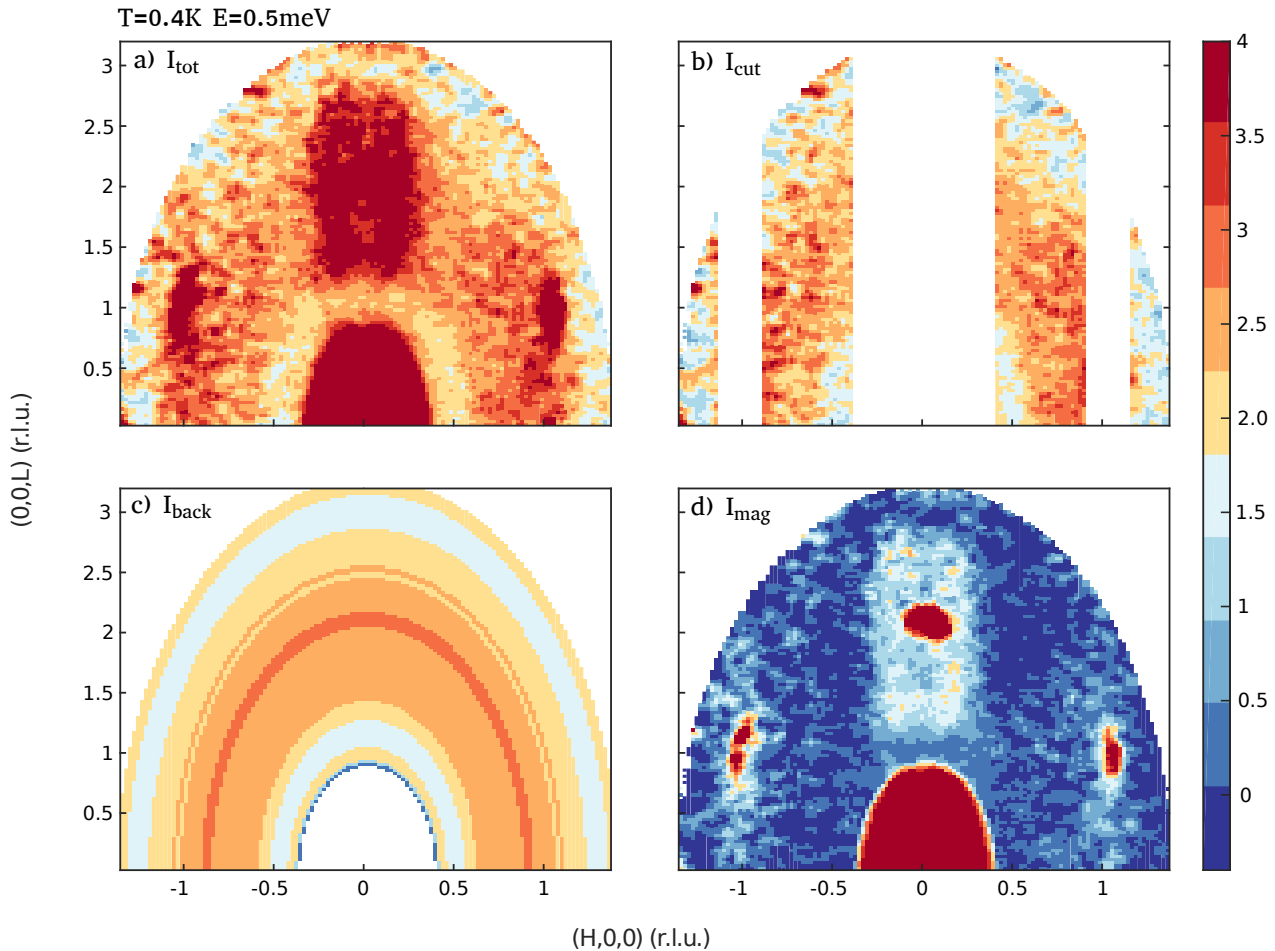}
\caption{The figure demonstrates the method used to determine the background. (a) illustrates the smoothed intensity. (b) the remaining data after removing the main magnetic signal, Bragg peaks and central beam. (c) is the background signal generated by finding the radial average of (b). (d) is the magnetic signal found by removing the background (c) from the smoothed data (a).}
\label{back}
\end{figure*}

\textit{Materials Preparation and Sample Mounting:}   Single crystals of CeRhSi$_{3}$ were synthesized using a flux technique.  Given the relatively small sample sizes for neutron scattering, an array of 2 g of single crystals were aligned on a series of Aluminum plates as displayed in Fig. \ref{sample_mount} $(c)$-$(d)$ with the rocking scan shown in Fig. \ref{sample_mount} $(a)$, indicating a mosaic of 4 degrees.  The individual single crystals were secured to the plates using hydrogen-free Fomblin oil and covered with Aluminum foil as shown in Fig. \ref{sample_mount} $(c-d)$.  The Aluminum plates were shaped in circles with two pins so that sample could be aligned and swapped between the H0L and HHL scattering planes.  Given the sample mounting, our sample was on average centrosymmetric as the mounting does not distinguish between $\pm$ $c$ of the individual crystals.

Heat capacity measurements were performed on a 3.2~$\pm$~0.5~mg sample, using a Physical Property Measurement System (PPMS, Quantum Design) in a temperature range between 1.2 and 3~K. A relaxation method with a 2$\tau$ fitting procedure was used. The data
displayed in Fig. \ref{sample_mount} $(b)$ shows a transition near 2 K expected from magnetic ordering.  

\textit{Neutron Spectroscopy:} Attempts to search for a static magnetic Bragg peak using the D23 diffractometer (ILL, France) were not successful in resolving a temperature dependent signal from the background. This maybe expected based on the crystal field analysis discussed above which suggests a comparatively weak static moment in comparison to the inelastic scattering.  We therefore focussed our measurements on studying the dynamics with spectroscopy.  

Initial triple-axis measurements were carried out on the PANDA spectrometer (FRM2) where it was established that the magnetic scattering was highly extended in momentum space.  This was further confirmed by experiments on the SPINS spectrometer (NIST).  Given the need for measurements of a broad range in momentum space, the MACS cold triple-axis spectrometer (NIST) was utilized.   Cooled filters of Beryllium and Beryllium Oxide were placed before the monochromator and after the sample respectively.  With a fixed E$_{f}$=3.7 meV, this configuration afforded energy transfers up to 1.3 meV.  The use of a double filter configuration  was found necessary to remove higher energy neutrons from scattering off the monochromator onto the sample.  In such a situation, without the first Beryllium filter, it was found that the sample mount and Fomblin oil gave a large background making extraction of the magnetic signal difficult and unreliable.    

Motivated by the observation of a commensurate response which maybe indicative of ferromagnetic interactions, we measured the neutron inelastic response under a magnetic field.  We note that in YbRh$_{2}$Si$_{2}$, the incommensurate response was found to be strongly affected by an applied field.~\cite{Stock12:109_2}  Measurements with a magnetic field were done using the IN12 cold triple-axis spectrometer at the ILL with E$_{f}$=3.5 meV and using a 6 T cryomagnet with a He-3 insert so that temperatures as low as 0.5 K could be reached.

\subsection{Background Subtraction on MACS}

Given our complex sample mounting scheme shown in Fig. \ref{sample_mount}, background scattering of neutrons from the Fomblin grease and also the Aluminum sample holder were an issue.   We outline here how the magnetic inelastic signal was isolated from the background using the detector arrangement on the MACS cold triple-axis spectrometer (NIST). The wide coverage of MACS allowed for a simultaneous measurement of both the magnetic inelastic signal and also a large background region where a comparatively strongly temperature dependent magnetic signal was not observed. 

The methodology used to calculate the background for any given temperature and energy is illustrated in Fig.~\ref{back}. Fig.~\ref{back} $(a)$ shows the smoothed intensity including background and magnetic scattering adding equivalent data at $\pm$ L positions. The intensity consists of a main magnetic region surrounded on both sides by a weaker scattering signal independent of sample rotation angle which appears as a ring of intensity in $|\vec{Q}|$. This background also includes Bragg peaks that occur near the (1, 0, 1) and (-1, 0, 1) positions. The relative strength of this background varied with temperature and energy. Therefore, the background needed to be approximated at each temperature and energy.

To subtract this background, for a given temperature and also energy transfer, we have removed strips that contained the magnetic signal and the Bragg peaks, as shown in Fig.~\ref{back} $(b)$. The remaining data was then used to determine a radially averaged background, as shown in Fig.~\ref{back} $(c)$. This averaging did not consider small angle scattering from the main beam indicated by the bright region about the origin in Fig. \ref{back}. Subtracting Fig.~\ref{back} $(c)$ from Fig.~\ref{back} $(a)$ returns the magnetic inelastic signal shown in Fig.~\ref{back} $(d)$.  A similar background subtraction procedure has been applied in a recent study on MACS investigating CeCo(In$_{1-x}$Hg$_{x}$)$_{5}$ ($x$=0.01).~\cite{Stock18:121}
 
\section{Experimental results}

\begin{figure}
\includegraphics[width=9.0cm] {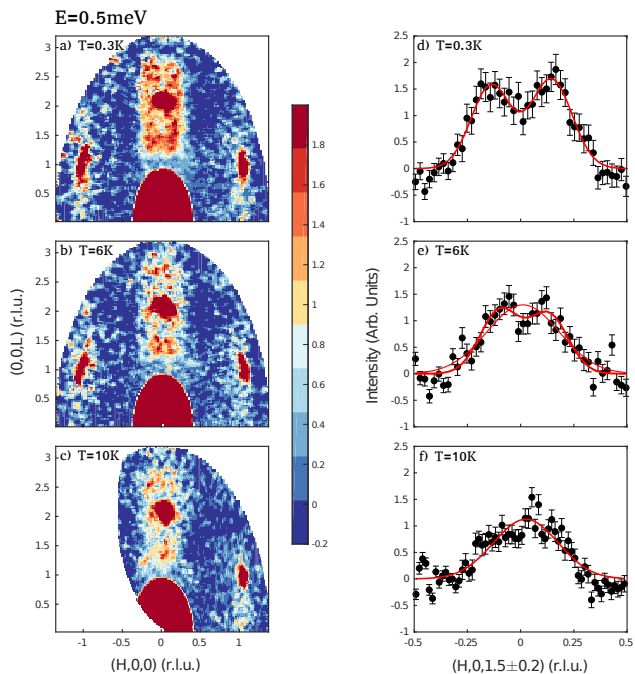}
\caption{\label{T_depend} Constant E=0.5 meV slices taken on MACS at $(a)$ T=0.3 K, $(b)$ 6 K, and $(c)$ 10 K.  Slices along the (H, 0, 1.5 $\pm$ 0.2) direction are shown in panels $(d)$-$(f)$.  The lines are fits to Gaussians with panel $(d)$ illustrating two symmetrically displaced peaks and $(e)$ showing a fit to two peaks and also a single peak.  At high temperatures of 10 K, $(f)$, the correlated scattering is well described by a single peak centered at the H=0 position.}
\end{figure}

Having discussed the single-ion properties of CeRhSi$_{3}$ and the experimental details, we discuss the results obtained for the dynamic neutron response.  We first show the temperature dependence of the correlated low-energy magnetic scattering.  Motivated by previous neutron scattering reports of a weak low temperature magnetic Bragg peak at ($\sim$ 0.2, 0, 0.5)~\cite{Aso07:310}, we focussed our measurements in the (H, 0, L) scattering plane.  Figure \ref{T_depend} shows the temperature dependence of the correlated magnetic scattering near $\vec{Q}$=(0,0,2) taken on MACS with the sample oriented in the (H0L) plane and with the background subtracted following the methodology discussed above.  Figure \ref{T_depend} $(a-c)$ illustrates a low temperature constant E=0.5 meV slices at T=0.3 K, 6 K, and 10 K.  The results show enhanced but momentum broadened scattering near $\vec{Q}$=(0,0,2) for these three temperatures.  We note that the commensurate scattering at the $\vec{Q}$=(0,0,2) is strongly contaminated by nuclear elastic scattering given that (0,0,2) is an allowed crystallographic Bragg peak.   The background corrected scattering near (0,0,2) in Fig. \ref{T_depend} $(a)$ consists of both a sharp component originating from scattering from the nuclear peak and also scattering elongated along the L direction and extending over the entire Brillouin zone.  The momentum broadened rods of scattering are elongated along L indicative of weak correlations along $c$.   With increasing temperature, shown in Figs. \ref{T_depend} $(b)$ and $(c)$, the scattering decreases in intensity, confirming the magnetic origin.

\begin{figure}
\includegraphics[width=7.5cm] {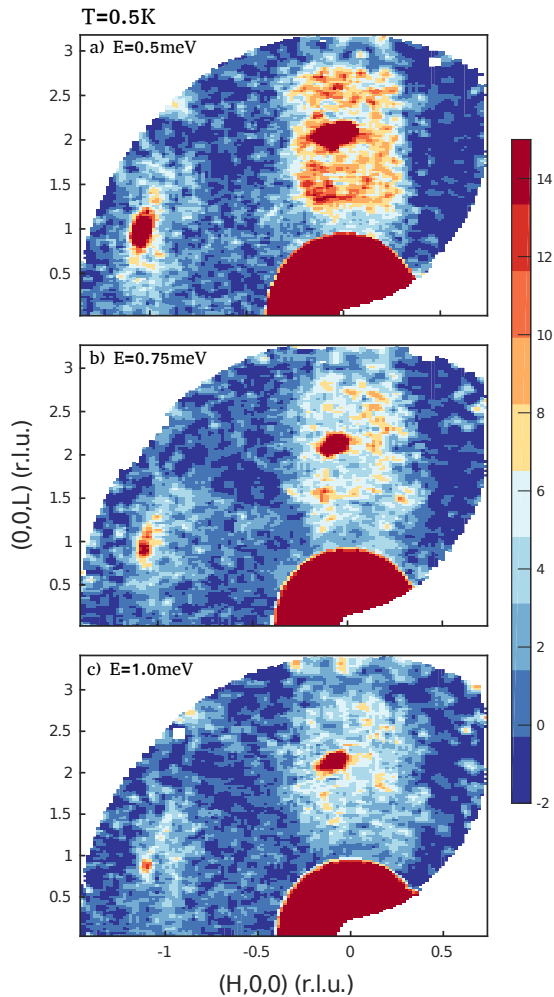}
\caption{\label{E_depend} T=0.5 K constant energy cuts at $(a)$ E=0.5 meV, $(b)$ 0.75 meV, and $(c)$ 1.0 meV.  The correlated magnetic scattering weakens and also shifts spectral weight to the commensurate position with increasing energy transfer.}
\end{figure}

Figure \ref{T_depend} panels $(d-f)$ illustrates cuts through these rods of scattering near $\vec{Q}$=(0, 0, 2) integrating along L in the (H, 0, 1.5 $\pm$ 0.2) direction.  At low temperatures (T=0.3 K) shown in Fig. \ref{T_depend} $(d)$, two clear peaks can be seen symmetrically displaced along H located at H$_{0}$=0.14 $\pm$ 0.03.  On heating, the structure disappears until 10 K where the scattering is well described by a commensurate peak centered at H=0.  The low-energy magnetic scattering in CeRhSi$_{3}$ therefore crosses over from incommensurate scattering along H to a commensurate response with increasing temperature.

\begin{figure}
\includegraphics[width=8.6cm] {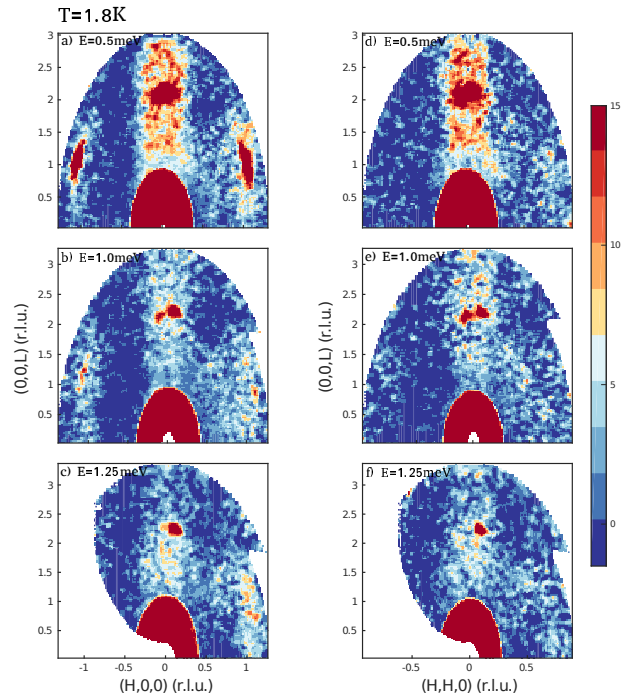}
\caption{\label{plane_compare} A comparison of the magnetic scattering in the (H, 0, L) (panels $a-c$) and (H, H, L) (panels $d-f)$) scattering planes.  }
\end{figure}

Having established the presence of momentum broadened correlations along L which weaken with increasing temperature as expected for magnetic scattering, we now show the energy dependence at low temperatures.  Figure \ref{E_depend} shows a series of constant energy slices taken at T=0.5 K using the MACS spectrometer.  At low-energy transfers of E=0.5 meV (panel $a$), extended scattering along L which is peaked near H=0.14 $\pm$ 0.03 r.l.u. is found.  However, as shown in Figs. \ref{E_depend} $(b)$ and $(c)$, on increasing the energy transfer the scattering becomes weaker and also centered around the commensurate $\vec{Q}$=(0,0,2) position.  For energy transfers at E=1.0 meV, shown in Fig. \ref{E_depend} $(c)$, the scattering remains highly extended in momentum yet more centered at the commensurate position.  The magnetic scattering therefore crosses over from an incommensurate response to a commensurate response through either increasing the energy transfer, or temperature.

The experimental data has primarily been taken in the (H,0,L) plane motivated by the discovery of a weak magnetic Bragg peak in this scattering plane.  However, in most tetragonal heavy fermion materials, like Ce(Rh,Co)In$_{5}$ family, the scattering is located near the antiferromagnetic (0.5, 0,5) position.~\cite{Stock15:114,Stock12:109,Stock12:109_2}  In Figure \ref{plane_compare}, the neutron response in both the (H,0,L) and (H,H,L) scattering planes is compared at a series of energy transfers.  The scattering in both planes is qualitatively similar with the magnetic scattering being highly extended along the L direction.  At low temperatures and energy transfers, in the (H, H, L) plane as illustrated in Fig. \ref{plane_compare} $(d)$, the scattering is also suggestive of being extended along the [1,$\overline{1}$,0] direction.  With increasing energy transfer, the scattering broadens and weakens considerably near the (0,0,2) position in both scattering planes.

\begin{figure}
\includegraphics[width=8.2cm] {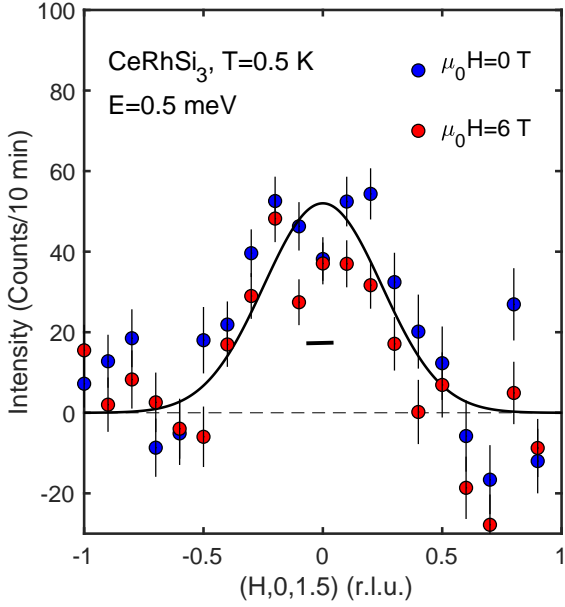}
\caption{\label{mag_figure}  Constant E=0.5 meV scans at 0 and 6 T taken on the IN12 spectrometer (ILL) with E$_{f}$=3.5 meV.  A temperature independent background has been subtracted from the data.  No strong or significant response of the magnetic fluctuations to field is observed.}
\end{figure}

The presence of magnetic scattering near the commensurate (002) position is suggestive of an underlying ferromagnetic response as observed in YbRh$_{2}$Si$_{2}$.~\cite{Stock12:109}  While ferromagnetism in Ce based compounds is rare, it has been reported in CeRuPO~\cite{Krellner07:76} and also CeSb$_{2}$~\cite{Canfield91:70}.  However, given the body centered nature of the crystallographic structure, it is also consistent with an antiferromagnetic interaction between nearest Ce$^{3+}$ neighbors along $c$.  To test for this hypothesis, we have applied a vertical magnetic field in order to check for a strong response of the magnetic fluctuations.  These experiments were performed on IN12 and are summarized in Fig. \ref{mag_figure} comparing a constant (H, 0, 1.5) scan at 0 and 6 T at low temperatures.  While the data is statistics limited, we do not observe a strong or significant response of magnetic fluctuations to a magnetic field aligned within the $a-b$ plane.   As an additional check for a possible ferromagnetic response, the elastic (0,0,2) nuclear Bragg peak was also studied as a function of magnetic field with no measurable change observed.  Based on this and the lack of a strong change with applied field in the inelastic channel, we conclude that the underlying interaction is dominated by antiferromagnetic interactions.

\section{Heuristic description}

Having presented the experimental results, we now attempt to understand the interplay between the commensurate magnetic fluctuations at high temperatures and energies and the crossover at low energies and temperatures to an incommensurate response.  We first discuss the results in terms of over damped $j_{eff}=1/2$ spin waves and show that the local crystal field symmetry is not consistent with such a description.  We then discuss the fluctuations in terms of weakly coupled Kramers doublets through the use of the Random Phase Approximation (RPA).  We find that this later approach allows us to reproduce, qualitatively, the interplay between commensurate and incommensurate responses. 

\subsection{$j_{eff}=1/2$ ground state and excitations?}

The crystal field scheme in Fig. \ref{structure_cef} shows that the ground state doublet is separated from the next higher energy doublet by $\sim$ 19 meV = 220 K.  Given the energy scale separating these two excitations, at low temperatures we would expect this doublet to be well separated from the higher energy and then to possibly behave as a $j_{eff}=1/2$ magnet similar to the spin-orbit split levels in Co$^{2+}$~\cite{Sarte18:98} or $4d$ or $5d$ transition metal ions.~\cite{Stamokostas18:97,Sarte18:98,Cowley13:88,Sakurai68:167,Kanamori57:17_1,Kanamori57:17_2}  In this section, we investigate this point by studying the lowest energy doublets and the angular momentum operators in this two dimensional subspace.

To investigate this point, we have transformed the coordinates of the angular momentum operators from a $|j,m\rangle$ basis to the a basis corresponding to the eigenstates of the the crystal field Hamiltonian $\mathcal{H}_{CEF}$ (denoted as the $|CEF\rangle$ basis).  This corresponds to writing a transformation matrix with the columns the eigenstates in the $|j,m\rangle$ basis as follows,

\begin{eqnarray}
\mathcal{C} = \left[ {\begin{array}{cccccc}
	0.605 	& 0  		& 0 		& 0 		& 0.796	& 0 \\
	0 		& 0.796  	& 0 		& 0 		& 0 		& -0.605  \\
	0 		& 0  		& 1.000 	& 0 		& 0 		& 0  \\
	0 		& 0  		& 0 		& 1.000 	& 0		& 0 \\
	-0.796 	& 0  		& 0 		& 0 		& 0.605	& 0  \\
	0 		& -0.605  	& 0 		& 0 		& 0 		& -0.796  \\
	\end{array} } \right] \nonumber.
\end{eqnarray}

\noindent Using this matrix, we can then transform any operator (denoted as $A$) from the $|j,m\rangle$ to the $|CEF\rangle$ basis and vice versa using

\begin{equation}
A_{|{CEF}\rangle} = \mathcal{C}^{-1}A_{|j,m\rangle}\mathcal{C}.
\label{eq:transform}
\end{equation} 

\noindent  $\mathcal{H}_{CEF}$ after this transformation takes the following diagonal form with the elements being the energy eigenvalues,

\begin{eqnarray}
\mathcal{H}^{CEF}_{|{CEF}\rangle} = \mathcal{C}^{-1}\mathcal{H}^{CEF}_{|j,m\rangle}\mathcal{C} = \nonumber \\
\left[ {\begin{array}{cccccc}
	-13.97 	& 0  		& 0 		& 0 		& 0		& 0 \\
	0 		& -13.97  	& 0 		& 0 		& 0 		& 0  \\
	0 		& 0  		& 5.16 	& 0 		& 0 		& 0  \\
	0 		& 0  		& 0 		& 5.16 	& 0		& 0 \\
	0 		& 0  		& 0 		& 0 		& 8.82	& 0  \\
	0 		& 0 	  	& 0 		& 0 		& 0 		& 8.82  \\
	\end{array} } \right]. \nonumber
\end{eqnarray}

\noindent  Using the transformation matrix $T$, we can transform the angular momentum operators between the different bases, for example with $J^{z}$ taking the following form,

\begin{eqnarray}
J^{z}_{|{CEF}\rangle} = \mathcal{C}^{-1}J^{z}_{|j,m\rangle}\mathcal{C} = \nonumber \\ 
\left[ {\begin{array}{cc|cccc}
	\textcolor{model7}{-0.034} 	& \textcolor{model7}{0}  		& 0 		& 0 		& 1.928	& 0 \\
	\textcolor{model7}{0} 		& \textcolor{model7}{0.034}  	& 0 		& 0 		& 0 		& -1.928  \\
	\hline
	0 		& 0  		& 0.5 	& 0 		& 0 		& 0  \\
	0 		& 0  		& 0 		& -0.5 	& 0		& 0 \\
	1.928	& 0  		& 0 		& 0 		& 1.034	& 0  \\
	0 		& -1.928 	& 0 		& 0 		& 0 		& -1.034  \\
	\end{array} } \right]. \nonumber
\end{eqnarray}

\noindent 	To determine if the ground state doublet can be projected onto a $j_{eff}$=1/2 angular momentum operator, we consider the upper $2 \times 2$ matrix (denoted in \textcolor{model7} {blue} ) which operates on this subspace.  Calculating this for the $\tilde{A}^{x}, \tilde{A}^{y},$ and $\tilde{A}^{z}$ operators we get, 

\begin{eqnarray}	
\tilde{A}^{x} = \left[ {\begin{array}{cc}
	0 		& 1.078  		\\
	1.078 	& 0  	 \\
	\end{array} } \right] \nonumber ,
\end{eqnarray}

\begin{eqnarray}
\tilde{A}^{y} = \left[ {\begin{array}{cc}
	0 		& -1.078i  		\\
	1.078i 	& 0  	 \\
	\end{array} } \right] \nonumber ,
\end{eqnarray}

\begin{eqnarray}
\tilde{A}^{z} = \left[ {\begin{array}{cc}
	-0.034 	& 0  		\\
	0 		& 0.034  	 \\
	\end{array} } \right] \nonumber ,
\end{eqnarray}

\noindent These $2 \times 2$ matrices do not follow the canonical commutation relations summarized by $\vec{J} \times \vec{J}=i\vec{J}$ as satisfied by the Pauli spin matrices which belong to the SU(2) group.  We therefore conclude that the ground state crystal field doublet of CeRhSi$_{3}$, while isolated at low temperatures from higher energy crystal field doublets, cannot be projected onto a $j_{eff}$=1/2 angular momentum operator.   Therefore, $\tilde{A}\neq \alpha J$, where $J$ is an angular momentum operator and $\alpha$ is a scaler projection factor. From a symmetry perspective, this result is not surprising given the highly anisotropic crystalline electric field.   We will discuss this result below in parametrizing the low-energy and temperature dependent spin fluctuations in CeRhSi$_{3}$.

\subsection{Correlated relaxing Kondo sites}

The magnetic scattering illustrated in Figs. \ref{T_depend} and \ref{E_depend} is highly extended both in momentum and energy, displaying little $\vec{Q}-\omega$ structure like that reported in other more localized Ce$^{3+}$ based systems such as CeRhIn$_{5}$~\cite{Stock15:114,Das14:113} or discussed in the context of temporally sharp spin excitations in CeCoIn$_{5}$~\cite{Song16:7}.  Therefore, parameterization of the excitations involving crystal fields or well defined harmonic magnons is not appropriate to describe the low-energy magnetic dynamics in CeRhSi$_{3}$ at ambient pressures.  Furthermore, the crystal field analysis discussed above illustrates that the ground state cannot be projected onto a $j_{eff}=\frac{1}{2}$ ground state and therefore damped spin wave theory involving Heisenberg coupling of $j_{eff}=\frac{1}{2}$ spins cannot be applied. Any heuristic model must be able to describe the temperature and energy dependence of the magnetic excitations discussed in the previous sections.  In particular, the results outlined above indicate two competing effects with one described by a commensurate wavevector near $\vec{Q}$=(0,0,2) and another associated with the low temperature incommensurate wave order appearing near H~$\sim$ 0.2.  In this section, we investigate whether a weakly correlated lattice of Kondo sites can consistently describe the commensurate scattering and the extended nature of it in momentum.  We emphasize that this model is not unique and similar incommensurate to commensurate magnetic responses have been observed in a number of materials notably in, for example, the cuprates and pnictides with models based on band structure or stripes.  

The magnetic neutron cross section is proportional to $S(\vec{Q},\omega)$ which is related to the susceptibility by,

\begin{eqnarray}
S(\vec{Q},\omega) \propto [n(\omega)+1] \chi '' (\vec{Q},\omega) \nonumber
\end{eqnarray}

\begin{figure}
\includegraphics[width=9.3cm] {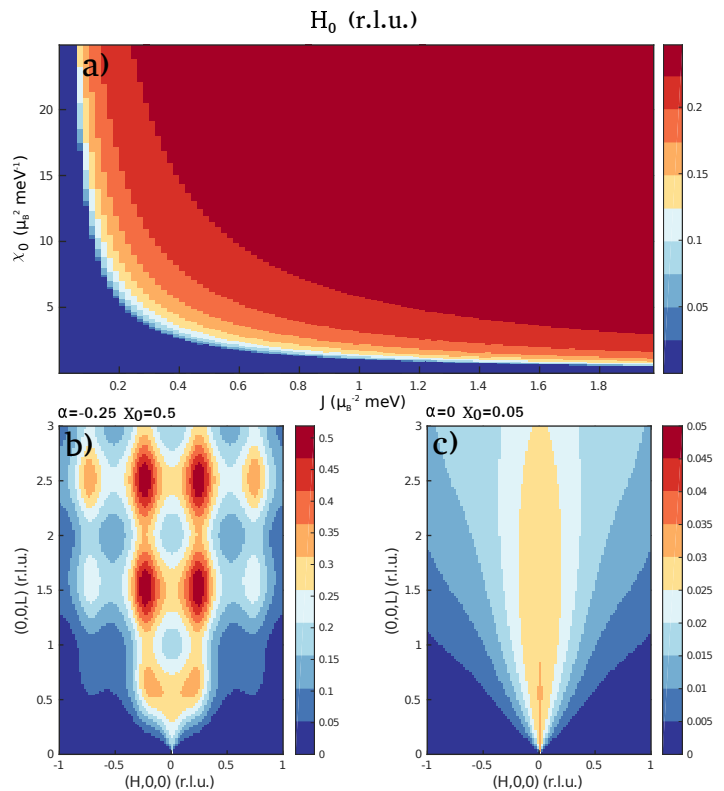}
\caption{\label{model_figure} $(a)$ The location in H of the maximum scattering intensity as a function of $X_{0}$ and exchange constant $J$ between different Ce$^{3+}$ ions.  $(b)$ A simulated constant E=0.5 meV slice taken with large $X_{0}$ and interlayer correlations $\alpha$.  $(c)$ A simulated constant energy slice with weak $X_{0}$ and interlayer correlations.  The calculations were done with $\Gamma$=0.5 meV.}
\end{figure}

\noindent where $[n(\omega)+1]$ is the Bose factor and $\chi '' (\vec{Q},\omega)$ is the imaginary part of the susceptibility.  To model the energy broadened commensurate component, we note that the extended nature of the neutron cross section in momentum indicates weak correlations between Ce$^{3+}$ sites.  We therefore consider a single-site susceptibility describing fluctuations within the ground state of the localized Kramers doublet with a characteristic energy scale $\Gamma$ set by the relaxational timescale $\tau$ via $\Gamma \sim \frac{1}{\tau}$.  This results in a single site susceptibility given by,

\begin{eqnarray}
\chi_{0}(\omega)={{X_{0} \Gamma} \over {\Gamma -i \omega}}, \nonumber
 \end{eqnarray}

\noindent  where $X_{0}$ is the temperature dependent single site static susceptibility.  Similar single-site susceptibilities have been used to describe the paramagnetic scattering in Uranium based heavy fermions~\cite{Broholm87:58,Stock11:107,Schroder98:80} and also in the paramagnetic phase of the cuprates~\cite{Yamani15:91,Stock08:77}.   In this case, given that the dominant single-ion cross section discussed above is for transitions between members of the ground state doublet, this susceptibility can be interpreted physically as transitions between the members of the lowest energy Kramers doublet with a timescale set by $\tau \propto {1 \over \Gamma}$.  We therefore term this single-ion response as a relaxing Kramers doublet.

To parameterize the coupling of these relaxing Kramers doublets on different Ce$^{3+}$ sites, we apply the random phase approximation (RPA) which gives a final susceptibility with the following form,

\begin{eqnarray}
\chi(\vec{Q},\omega)= {\chi_{0} (\omega)  \over {1-\chi_{0}(\omega)J(\vec{Q})}}. \nonumber
 \end{eqnarray}

\noindent Where $J(\vec{Q})$ is the Fourier transform of the coupling between the localized magnetc sites each consisting of a low energy two level energy scheme. We consider the following heuristic form for a two dimensional magnet with only nearest neighbor coupling along the $a$ and $b$ axes for simplicity,  

\begin{eqnarray}
J(\vec{Q})= \sum_{r} J_{r} e^{i\vec{Q}\cdot\vec{r}}\\ \nonumber
=2J_{RKKY} \left[ \cos(2\pi H)+\cos(2\pi K) \right ]
 \end{eqnarray}
 
\noindent  Putting this all together to calculate the imaginary part of the susceptibility proportional to the neutron cross section gives,
 
 \begin{eqnarray}
\chi''(\vec{Q},\omega)= {{X_{0} \Gamma \omega} \over {\omega^{2}+\Gamma^{2}(1-X_{0}J(\vec{Q}))^{2}}}.
 \end{eqnarray}
 
 \noindent This relation highlights the fact that the cross section is peaked in momentum when $1-X_{0}J(\vec{Q})$ is a minimum.  We note that this form of $\chi''$ is an odd function in energy required for detailed balance.~\cite{Shirane}  Given the scattering is confined along the (0,0,L) direction and extended along L indicative of short-range correlations, we have considered the cross section from in-plane fluctuations and weakly correlated Ce$^{3+}$ spins along $c$.  The neutron cross section can therefore be written as,
 
 \begin{eqnarray}
S(\vec{Q},\omega) \propto [n(\omega)+1] \chi '' (\vec{Q},\omega) \left[1-\left({Q_{ab} \over Q}\right)^{2} \right ] ... \\ \nonumber
\times \left[1+ \alpha \cos (\vec{Q} \cdot \vec{c}) \right].
\end{eqnarray}

\noindent The parameter $\alpha$ denotes the strength of nearest neighbor correlations along $c$.  We note the distinction between coupling, described by $J (\vec{Q})$, and correlations parameterized by the susceptibility and $\alpha$.

Fig. \ref{model_figure} $(a)$ illustrates the $X_{0}$ and $J$ phase diagram with the colors indicating where in $H$ the term in the susceptibility $\left(1-X_{0}J(\vec{Q})\right)$ is a minimum and therefore where the neutron scattering cross section is maximum.  The value of $\Gamma$ was chosen to be 0.5 meV to match the energy range were strong magnetic fluctuations are observed on MACS.  The plot is done for $X_{0}$ and $J$ and shows that for small values of the susceptibility, $X_{0}$, a maximum in the scattering cross section occurs at the commensurate value H$\sim$ 0 where for larger values of either $X_{0}$ or $J$, the cross section is maximum at H=0.25 when $2\pi H=1/2$.  This is further illustrated in Figs. \ref{model_figure} $(b,c)$ which plots constant energy slices for the two extreme cases.  Panel $(b)$ shows the scattering cross section for the case of both large susceptibility $X_{0}$ and also large correlations along $c$.  The intensity profiles display peaks displaced along H near H=0.25 and also L=0.5.  When the susceptibility $X_{0}$ is reduced, the cross section becomes peaked around the commensurate position as illustrated in panel $(c)$.

This simple and heuristic model of correlated and relaxing localized Kramers doublets captures the main qualitative results of our experiments.   The calculations with $\Gamma$=0.5 meV provide an estimate for the exchange constant of $\sim$ 1 meV (assuming $X_{0}\sim 1 \mu_{B}^{2}/meV$) to be compared with an estimate of $\sim$ 10 meV extracted from the Kondo temperature and specific heat coefficient.  Our choice of $X_{0}$ is consistent with $Q=0$ susceptibility measurements which indicate a low temperature susceptibility of order $\sim 10^{-2}$ emu/mol.~\cite{Muro07:76}  Our calculated value is sensitive to the choice of $\Gamma$ also $X_{0}$, but nevertheless is in reasonable agreement with the value estimated based on Ref. \onlinecite{Yang08:454}.

The experimental results shown above exhibit a trade off in intensity between incommensurate magnetic scattering peaked near ($\sim$ 0.2,0,1.5) and also the commensurate (0,0,2) position.  The maximum in the cross section shifts with increasing temperature or energy transfer.  In both cases, our model would imply that such a change in either energy or temperature coincides with a reduction in susceptibility $X_{0}$.  We note that neutron diffraction results have reported a Bragg peak near $\vec{q}_{0}$=($\sim$ 0.22, 0, 0.5).  The displacement of the wavevector from the 0.25 position along H may be accounted for through incorporating further neighbor exchange constants resulting in a more complex form of $J(\vec{Q})$ discussed above.

\section{Conclusions and Discussion}

The momentum and energy broadened response reported here for CeRhSi$_{3}$ differs from dispersing spin waves measured in insulating magnets and also metallic compounds with strongly localized moments.  In this context CeRhSi$_{3}$ differs from other non centrosymmetric systems like CePt$_{3}$Si which displays momentum dispersing spin wave excitations, albeit damped.~\cite{Fak08:78}   The energy and momentum broadened spin response at low energies also contrasts to the localized systems with the same crystal structures such as CeCoGe$_{3}$~\cite{Smidman13:88} and CeIrGe$_{3}$~\cite{Anand18:97} and is suggestive of a highly itinerant $f$-electrons.  This latter point is consistent with quantum oscillation measurements which fail to measure a divergence in effective mass in the magnetic phase as a function of pressure which would indicate a crossover to strongly localized moments.~\cite{Terashmia07:76}  

The results presented here also differ from muon results which suggest static magnetism on the MHz timescale.  We observe relaxational dynamics with a frequency scale on the order of THz, and this may indeed be outside the time window of muons.  The pressure dependence of the magnetic ordering~\cite{Egetenmeyer12:108} has been interpreted in terms of the Doniach phase diagram~\cite{Doniach77:91,Yang08:454} which captures the competition between RKKY driven antiferromagnetism and a Kondo screened state~\cite{Coleman83:28,Auerbach86:57,Lacroix79:20} separated by a critical exchange constant value.   Our results point to CeRhSi$_{3}$ being on the borderline between these different phases with relaxational dynamics reminiscent of a screened antiferromagnetic phase without long range order.  However, our results are consistent with a weak RKKY coupling between sites and also a growing susceptibility which drives the incommensurate order.  This might point to the existence of well defined spin excitations at low-energies in more localized variants such as CeCoGe$_{3}$ where much higher pressures ($\sim$ 4 GPa) are required to achieve superconductivity~\cite{Settai07:310} and where also thermodynamic measurements suggest a weaker Kondo effect than in CeRhSi$_{3}$~\cite{Muro98:67}. We would expect a decrease in J$_{RKKY}$ with pressure as Kondo screening becomes more prevalent.  This is consistent with thermoelectric measurements under pressure~\cite{Tannaka13:62} and also resistivity on CeRhSi$_{3}$.~\cite{Tchokonte01:117}

In summary, we have measured the low energy magnetic response in CeRhSi$_{3}$ at ambient pressures and at low temperatures.  A momentum and energy broadened response is observed which crosses over from commensurate at high energies to a incommensurate wave vector along $H$ at low energy transfers and low temperatures.  The response is not consistent with dispersing or damped spin excitations as observed in other Ce based non centrosymmetric superconductors.  We find a heuristic model based on weakly correlated relaxing and strongly Kondo screened localized Kramers doublets qualitatively accounts for this behavior.  We suggest that CeRhSi$_{3}$ is on the border between a Rudderman-Kittel-Yosida antiferromagnetic phase with strong and underdamped magnetic correlations and a fully Kondo screened phase where no correlations exist.  

The work here was funded by the EPSRC, the Carnegie Trust for the Universities of Scotland, and the STFC.  Access to MACS was provided by the Center for High Resolution Neutron Scattering, a partnership between the National Institute of Standards and Technology and the National Science Foundation under Agreement No. DMR-1508249.  Work at Brookhaven National Laboratory was supported by US DOE, Office of Science, Office of Basic Energy Sciences (DOE-BES) under Contract No. DE-SC0012704


%

\end{document}